\documentclass[preprint2]{aastex}

\usepackage{amsmath}
\usepackage{graphicx}
\usepackage{epsfig,psfrag,epic,eepic}
\usepackage{textcomp}
\usepackage{times}

\slugcomment{Accepted by Astrophysical Journal Letters}

\shorttitle{NIR Photometry of GJ~758~B}
\shortauthors{Janson et al.}

\begin{document}

\title{Near-Infrared Multi-Band Photometry of the Substellar Companion GJ~758~B\altaffilmark{*}}

\author{M. Janson\altaffilmark{1}, 
J. Carson\altaffilmark{2}, 
C. Thalmann\altaffilmark{3}, 
M.~W. McElwain\altaffilmark{4}, 
M. Goto\altaffilmark{3}, 
J. Crepp\altaffilmark{5}, 
J. Wisniewski\altaffilmark{6}, 
L. Abe\altaffilmark{7}, 
W. Brandner\altaffilmark{3}, 
A. Burrows\altaffilmark{4}, 
S. Egner\altaffilmark{8}, 
M. Feldt\altaffilmark{3}, 
C.~A. Grady\altaffilmark{9}, 
T. Golota\altaffilmark{8}, 
O. Guyon\altaffilmark{8}, 
J. Hashimoto\altaffilmark{10}, 
Y. Hayano\altaffilmark{8}, 
M. Hayashi\altaffilmark{8}, 
S. Hayashi\altaffilmark{8}, 
T. Henning\altaffilmark{3}, 
K.~W. Hodapp\altaffilmark{11}, 
M. Ishii\altaffilmark{8}, 
M. Iye\altaffilmark{10}, 
R. Kandori\altaffilmark{10}, 
G.~R. Knapp\altaffilmark{4}, 
T. Kudo\altaffilmark{10}, 
N. Kusakabe\altaffilmark{10}, 
M. Kuzuhara\altaffilmark{10,17}, 
T. Matsuo\altaffilmark{12}, 
S. Mayama\altaffilmark{19}, 
S. Miyama\altaffilmark{10}, 
J.-I. Morino\altaffilmark{10}, 
A. Moro-Mart\'in\altaffilmark{13}, 
T. Nishimura\altaffilmark{8}, 
T.-S. Pyo\altaffilmark{8}, 
E. Serabyn\altaffilmark{12}, 
H. Suto\altaffilmark{10}, 
R. Suzuki\altaffilmark{10}, 
M. Takami\altaffilmark{14}, 
N. Takato\altaffilmark{8}, 
H. Terada\altaffilmark{8}, 
B. Tofflemire\altaffilmark{6}, 
D. Tomono\altaffilmark{8}, 
E.~L. Turner\altaffilmark{4,18}, 
M. Watanabe\altaffilmark{15}, 
T. Yamada\altaffilmark{16}, 
H. Takami\altaffilmark{8}, 
T. Usuda\altaffilmark{8}, 
M. Tamura\altaffilmark{10}
}

\altaffiltext{*}{Based on data collected at Subaru Telescope, which is operated by the National Astronomical Observatory of Japan, on Gemini data under program GN-2010A-Q-23, and on Keck data under N068N2.}
\altaffiltext{1}{Univ. of Toronto, Canada; \texttt{janson@astro.utoronto.ca}}
\altaffiltext{2}{College of Charleston, Charleston, South Carolina, USA}
\altaffiltext{3}{Max Planck Institute for Astronomy, Heidelberg, Germany}
\altaffiltext{4}{Dep.\ of Astroph.\ Sciences, Princeton Univ.,\ Princeton, USA}
\altaffiltext{5}{California Institute of Technology, Pasadena, USA}
\altaffiltext{6}{University of Washington, Seattle, Washington, USA.}
\altaffiltext{7}{Laboratoire Hippolyte Fizeau, Nice, France}
\altaffiltext{8}{Subaru Telescope, Hilo, Hawai`i, USA}
\altaffiltext{9}{Eureka Scientific \& Goddard Space Flight Center, USA}
\altaffiltext{10}{National Astronomical Observatory of Japan, Tokyo, Japan}
\altaffiltext{11}{Inst.\ for Astron., University of Hawai`i, Hilo, Hawai`i, USA}
\altaffiltext{12}{Jet Propulsion Laboratory, Caltech, Pasadena, USA}
\altaffiltext{13}{Dep.\ of Astroph., CAB - CSIC/INTA, Madrid, Spain}
\altaffiltext{14}{Inst.\ of Astron.\ and Astroph., Academia Sinica, Taipei, Taiwan}
\altaffiltext{15}{Dep. of Cosmosciences, Hokkaido University, Sapporo, Japan}
\altaffiltext{16}{Astronomical Institute, Tohoku University, Sendai, Japan}
\altaffiltext{17}{University of Tokyo, Tokyo, Japan}
\altaffiltext{18}{Inst.\ for Ph. and Math. of the Universe, Univ. of Tokyo, Japan}
\altaffiltext{19}{Grad. Univ. for Adv. Studies, Kanagawa, Japan}

\begin{abstract}\noindent
GJ~758~B is a cold ($\sim$600K) companion to a Sun-like star at 29 AU projected separation, which was recently detected with high-contrast imaging. Here we present photometry of the companion in seven photometric bands from Subaru/HiCIAO, Gemini/NIRI and Keck/NIRC2, providing a rich sampling of the spectral energy distribution in the 1--5 $\mu$m wavelength range. A clear detection at 1.58$\mu$m combined with an upper limit at 1.69$\mu$m shows methane absorption in the atmosphere of the companion. The mass of the companion remains uncertain, but an updated age estimate indicates that the most likely mass range is $\sim$30--40 $M_{\rm jup}$. In addition, we present an updated astrometric analysis that imposes tighter constraints on GJ~758~B's orbit and identifies the proposed second candidate companion, ``GJ~758~C'', as a background star.
\end{abstract}


\section{Introduction}

In recent years, a number of high-contrast companions have been thermally imaged around nearby stars \citep[e.g.][] {marois2008, lagrange2010}. One interesting companion is GJ~758~B \citep[hereafter P1]{thalmann2009}. Its combination of a Sun-like parent star (spectral type G8V) only 15.5 pc away \citep{perryman1997}, the close proximity to the star (projected separation 29\,AU), and low surface temperature ($\sim$600 K) make it one of the most ``planet-like'' objects available for direct study, acting as a laboratory for current planet formation and evolution theories.

P1 presented two epochs of $H$-band imaging of GJ~758~B that allowed for proof of common proper motion and a first estimation of physical and orbital properties.  Furthermore, a candidate second companion, tentatively called ``GJ~758~C'', was found in one epoch. Recently, \citet{currie2010} published $L^\prime$-band data that confirmed the estimated temperature range of B but did not detect ``C''. Both publications deduced similar results for best-fit orbital parameters ($a \sim 50$ AU, $e \sim 0.7$). Both also noted that the mass range is $\sim$10--40 $M_{\rm jup}$ if the full range of main-sequence ages for a Sun-like star is considered. In this work, we present photometric coverage of GJ~758~B in the near infrared ($JHK_\mathrm{c}L^\prime M$ bands) with additional photometry in the methane-sensitive CH4S and CH4L narrow-band filters (see Table~\ref{t:results} for filter specifications), collected with high-contrast imaging techniques at Subaru/HiCIAO, Gemini/NIRI, and Keck/NIRC2. This is used for updating the estimations of physical and orbital parameters of GJ 758 B. We report the re-detection of ``GJ~758~C'' and the conclusive identification of it as a background star by proper motion.

\section{Observations and Data Reduction}

$H$-band observations of GJ~758~B were obtained with Subaru on November 11, 2009. making use of AO188 \citep{hayano2010} coupled with HiCIAO \citep{hodapp2008} as part of the SEEDS survey \citep{tamura2009}.  The field of view (FOV) was 20$\arcsec$$\times$20$\arcsec$ with a 0$\farcs$010 pixel scale.  The total integration time was 906 s.

Observations in $J$-, $H$-, CH4S-, CH4L-, and $K_{\rm c}$-band were obtained at Gemini North with the Altair AO system \citep{herriot2000} and NIRI \citep{hodapp2003}. The observations were carried out on 5 different nights between April 27 and May 08, 2010, with a 0$\farcs$022 pixel scale over a 22$\arcsec$$\times$22$\arcsec$ FOV. The total integration times were 900 s, 1800 s, 3810 s, 2220 s, and 2400 s, for the $J$-, $H$-, CH4S-, CH4L-, and $K_{\rm c}$- band, respectively.  The nearby star HD 226294 was observed before or after each ADI dataset in the same filter to be used as a photometric reference.

We imaged GJ~758 in the $L^{\prime}$- and $M_{\rm s}$-bands on August 6, 2010 using the Keck II AO system \citep{vandam2004} and NIRC2. The 0$\farcs$300 diameter coronagraphic mask was used for the $L^{\prime}$ observations to prevent saturation. Images without the mask were also taken, to calibrate the brightness of the star. For all observations, we avoided the bad quadrant in the NIRC2 detector and used the 768$\times$776 subarray mode with a 0$\farcs$010 plate scale. The integration times were 1800 s and 1995 s for $L^{\prime}$ and $M_{s}$, respectively.Three photometric standard stars (HD 162208, HD 161903, Gl 748) were observed in both bands.

\begin{figure*}[p]
\centering
\includegraphics[width=\linewidth]{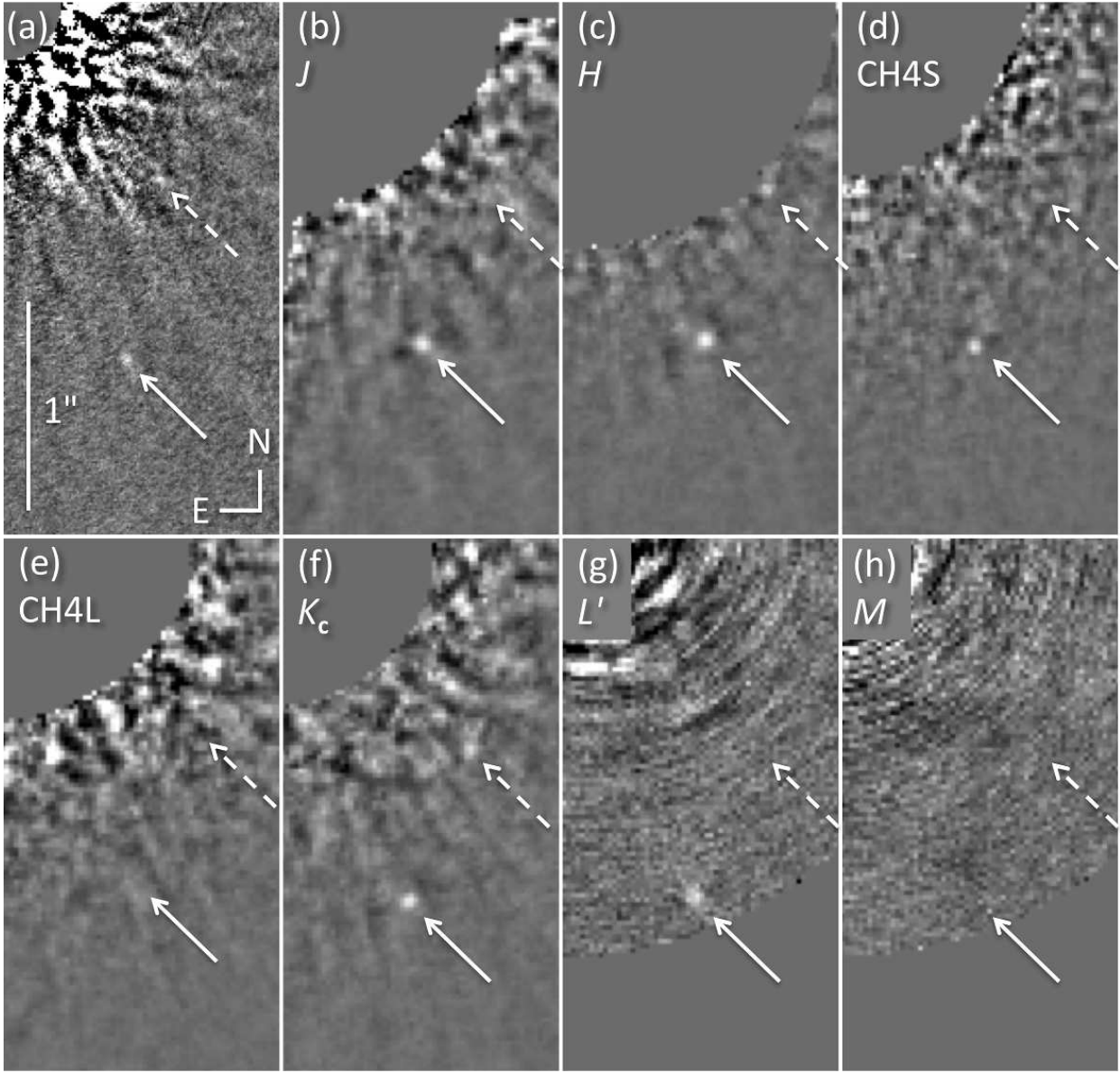}
\caption{High-contrast imaging of GJ~758~B with ADI/LOCI. In all panels, the star is located approximately in the upper left corner, North is up, and East is left.  The location of GJ~758~B is marked with a solid arrow, that of ``GJ~758~C'' with a dashed arrow. (\textbf{a}) HiCIAO $H$-band image (Nov 2009). GJ~758~B is clearly visible, whereas ``C'' is marginally detectable at $\sim$$3\sigma$. (\textbf{b--f}) Gemini/NIRI images in five filters (Apr/May 2010). (\textbf{g--h}) NIRC2 images in $L^\prime$- and $M_{\rm s}$-band (Aug 2010).}
\label{f:images}
\end{figure*}

\begin{figure*}[p]
\centering
\includegraphics[width=\linewidth]{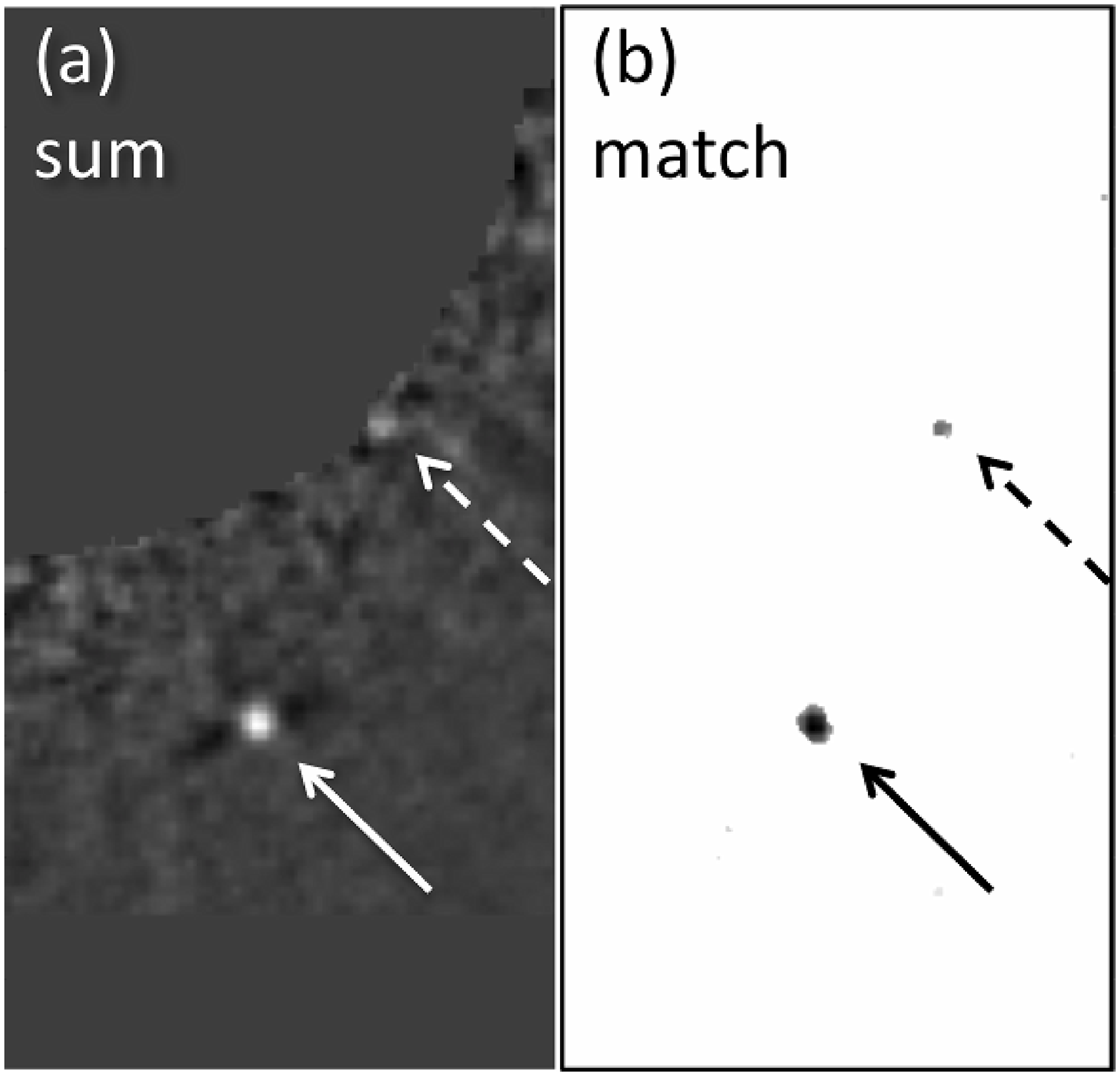}
\caption{Combination of Gemini images. (\textbf{a}) Co-add of $J$, $H$, $K_\mathrm{c}$, and CH4S, after normalizing the flux levels such that GJ~758~B appears approximately equally bright in each image.  The ``C'' source is detectable. (\textbf{b}) A map of the pixels that exhibit a S/N ratio of $\ge$+1 in all four images used for (a). Only GJ~758~B and ``C'' pass the test, indicating that they are both real on-sky objects.}
\label{f:ctest}
\end{figure*}

All data were taken using angular differential imaging (ADI, e.g. \citealt{marois2006}) with typical field rotations of $\sim$30$^{\rm o}$, and the images were reduced with the LOCI procedure \citep{lafreniere2007}, using an IDL implementation adapted for each of the instruments. For the HiCIAO data, the same procedure as described in P1 was used, including correction for the partial subtraction of companion flux during the LOCI reduction. The only practical differences for NIRI and NIRC2 were the image registration procedures, where the NIRI registration was performed on the basis of cross-correlation of the diffraction spiders, and NIRC2 registration was caclulated by centroiding on the stellar PSF core, which was non-saturated in the M-band data and non-saturated behind the semi-transparent mask in the $L^\prime$-band data. Also, for the NIRC2 data, high-pass filtering was applied at the same time as the subtraction of a radial profile from the stellar PSF in all images, to remove low-frequency spatial variations in the high thermal background of the $L^\prime$- and $M_{\rm s}$-band \citep[e.g.][]{janson2008}. Fluxes of all observed targets and standard stars were extracted using aperture photometry. The reduced images are shown in Fig. \ref{f:images} and Fig. \ref{f:ctest}.

\section{Photometric Analysis}

\subsection{Calibration}

Photometric calibration is of central importance for these data, due to the fact that the $JHK_{\rm s}$-photometry of GJ~758~A has been flagged as unreliable in 2MASS \citep{skrutskie2006}, and furthermore, for the broad-band filters it is always saturated in NIRI images, even for the shortest available integration times in subarray mode. Hence, we used the standard star for photometric calibration in all filters, with the exception of CH4S. Those data were taken on Apr 29 under photometrically unstable conditions, so that the standard star photometry couldn't be trusted. Instead, we calibrated the CH4S flux based on the fluxes of the background stars in the GJ~758 field in CH4L and CH4S, under the assumption that they are equally bright in both filters, which is reasonable since they are stars and cannot exhibit methane absorption. Four background stars were used with a dispersion of 0.19 mag. The brightness of the background stars in CH4L was calibrated against the standard star. The resulting absolute magnitudes of GJ~758~B are summarized in Table \ref{t:results}.

GJ~758~A has no published $L^\prime$-band or $M_{\rm s}$-band magnitudes, thus we use the standard stars to calibrate those values. All three standard stars give consistent results within 0.1 mag, yielding $M_{\rm L^\prime} = 3.7 \pm 0.1$ mag and $M_{\rm Ms} = 3.6 \pm 0.1$ mag. During the science exposures, GJ 758 A was unsaturated in the $M_{\rm s}$-band and so could be used to evaluate the brightness of GJ 758 B. In the case of $L^\prime$, the star was behind the semi-transparent mask, which makes it unreliable for photometry, hence for this case, standard star photometry was used directly to calibrate the brightness of GJ~758~B. The results are in Table \ref{t:results}. There are two cases where special circumstances apply: For CH4L, the flux at the position of the companion is visually unconvincing as a point source, and more reminiscent of residual noise from the stellar PSF. It is therefore possible that GJ~758~B is systematically overestimated at this wavelength, such that the quoted CH4L flux should rather be regarded as an upper limit rather than the true brightness. In the case of $M_{\rm s}$-band, there is no visually clear signature of the companion, but aperture photometry was performed on the same location as the $L^\prime$-band detection. This yielded an excess flux at the $\sim$2$\sigma$ level, hence this is a plausible flux of the companion, since its position is known a priori. Aside from CH4L and $M_{\rm s}$, all detections are $\sim$10$\sigma$ confidence. For NIRI, the photometric error that we calculate is dominated by an uncertainty in the linearity behaviour of the detector at low counts/short integrations, which we estimate could impose an error of at most 20\%.

\subsection{Interpretation}

\begin{figure*}[p]
\centering
\includegraphics[width=\linewidth]{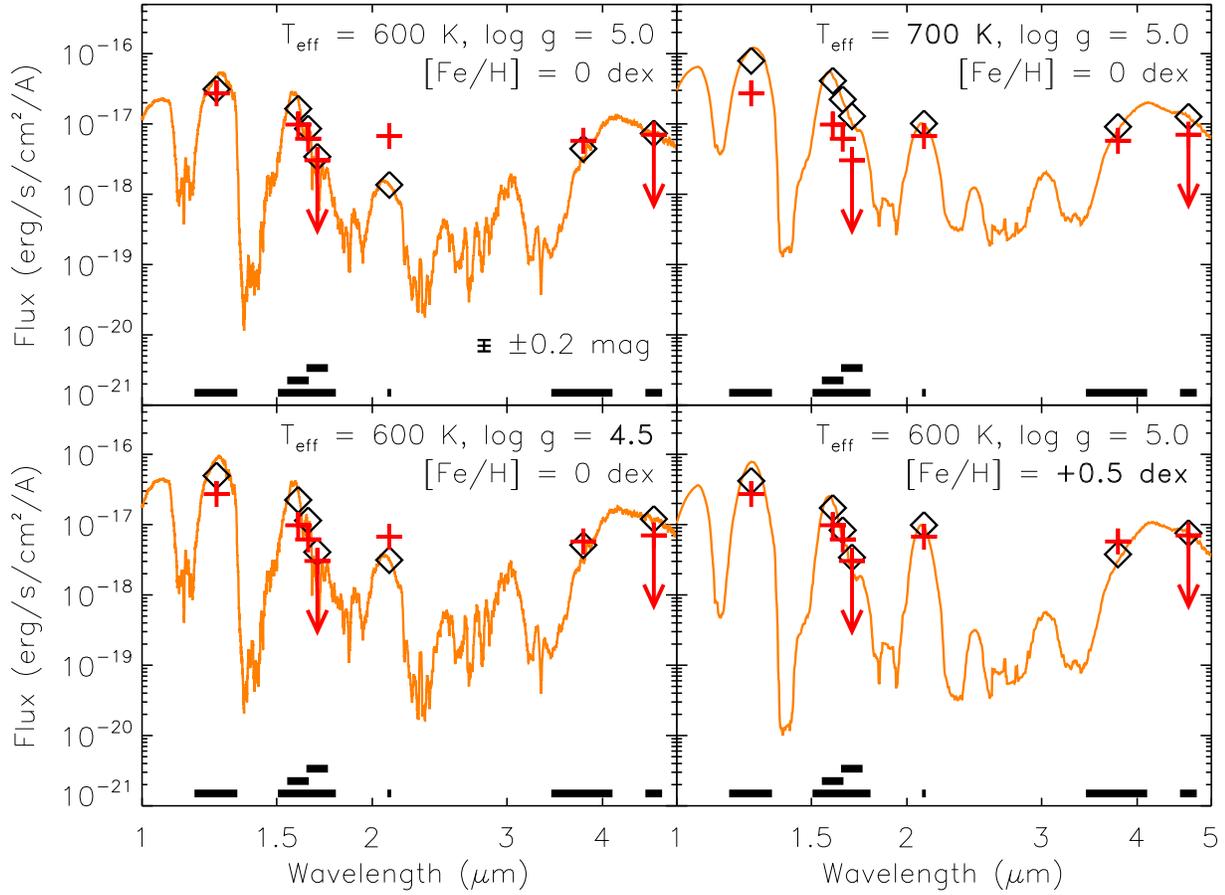}
\caption{Photometric analysis of GJ~758~B. The red plus signs show the measured flux values normalized to a distance of 10 pc for the seven filter bands, whose wavelength domains are marked with black bars. The orange curves represent model spectra for different assumptions of $T_\mathrm{eff}$, $\log g$, and metallicity, and the black diamonds are the resulting flux levels in the filter bands.}
\label{f:spectra}
\end{figure*}

GJ~758~B exhibits clear methane absorption. This is generally expected for objects in the late T-type range, although to a decreased extent if the atmosphere is in chemical non-equilibrium \citep[e.g.][]{burgasser2006,fortney2008}. GJ 758 A has a metallicity of approximately +0.2 dex \citep[e.g.][]{holmberg2009, kospal2009}. Assuming that the companion has the same composition, this super-solar metallicity is likely at least partly responsible for the high flux in $K_\mathrm{c}$. Since the commonly used COND models \citep{allard2001, baraffe2003} do not contain any non-solar abundances at the relevant temperature range, we do not use them for model comparison. Instead, we use models based on \citet{burrows2006} extended to colder temperatures (Hubeny \& Burrows in prep.). With our new adopted age range discussed below, the feasible values of $\log g$ for the companion are all close to 5.0, hence we use this as a baseline value in our fitting. The comparison was done by fitting model spectra to the flux densities of the photometric points, which were derived from the magnitude values using the NIRI and NIRC2 filter transmission curves. All fluxes were normalized to a distance of 10 pc. The best-fit value was acquired using the minimized $\chi^2$ in the photometric bands as the quality metric. The models cover temperatures of 500--700 K in steps of 50 K, $\log g$ of 4.0--5.0 in steps of 0.5 dex, and metallicities of +0.0 and +0.5 dex (note that the full grid is not covered -- in particular, no super-solar abundace models exist below 600 K). The best-fit temperature of this procedure is 600 K, in agreement with previous results. A metallicity of +0.5 dex fits better than +0.0 dex, which is consistent with the metal enrichment expected in the system. However, since surface gravity and metallicity are largely degenerate for the wavelengths covered by our study and the models are very uncertain, we do not attempt to optimize the fit with respect to either of these quantities. Indeed, the models are known to be unable to reproduce various features of other cool companions to stars \citep[e.g. the HR~8799 planets, see][]{marois2008, janson2010, hinz2010, bowler2010}, so deriving absolute values from model comparison is of limited relevance. Four examples of model fits are shown in Fig. \ref{f:spectra}.

We also make a comparison with field brown dwarfs from \citet{leggett2010}. For this purpose, we use the J-band absolute brightness $M_{\rm J} = 17.58 \pm 0.20$ mag, and the color $J - H = 0.58 \pm 0.28$, mag, since this should be relatively insensitive to metallicity and gravity. The companion fits very well to the temperature sequence of the \citet{leggett2010} field dwars in an HR-diagram, among the latest-type brown dwarfs (T8--T9). The $J - H$ color alone with uncertainties places the companion firmly beyond T5. A T8--T9 spectral type fits well to the temperature of $\sim$600 K derived from model comparison.

In P1, we adopted an age range of 0.7--8.7 Gyr, on the basis of a variety of age indicators. The 0.7 Gyr age was based on isochronal fitting by \citet{takeda2007}. However, other isochronal fits have been performed yielding substantially higher ages of several Gyr \citep[e.g.][]{valenti2005, holmberg2009}. This indicates that isochronal fitting is inadequate for dating GJ 758, hence we revise our age estimates by removing the isochrone method, which yields a residual age range of $\sim$5--9 Gyr (keeping estimates based on activity and rotation). As an additional age determination test, we have analyzed a high-resolution ($R$$\sim$31,500) spectrum of GJ~758 from the Apache Point Observatory 3.5m. We detect no clear evidence of the Li I 6708 \AA\ doublet at an upper limit of $\sim$4 m\AA\ equivalent width. If Li had been detected, it would have indicated a young age, but the non-detection does not provide a strong lower limit on the age, as it only suggests that GJ~758 should be older than the $\sim$100 Myr old Pleiades cluster \citep{maldonado2010}. We reiterate that age determination of main-sequence Sun-like stars is highly uncertain, and that the adopted age range of $\sim$5--9 Gyr should not be considered definitive, but merely represents the range of mean ages from methods that are considered sufficiently reliable. The corresponding range of companion masses from evolutionary models is $\sim$30--40 $M_{\rm jup}$. The most promising avenue for getting a reliable age estimate of GJ 758, and thus a better constraint on the mass of its companion, is likely asteroseismology.

\section{Astrometric Analysis}

\subsection{Data points}

In P1, we demonstrated the common proper motion of GJ~758 and GJ~758~B on the basis of two epochs of observation, May 3, 2009 (E1) and August 6, 2009 (E2).  In this work, we include three additional epochs in our analysis:  Our data from November 1, 2009 (E3) and April 29, 2010 (E4), as well as the May 27, 2010 data point from \citet{currie2010} (E5).  The Keck/NIRC2 data from August 2010 (E6) have a narrow field of view excluding the 5--7 known background stars that were used to fine-tune the pixel scales and rotation angles in our previous observation. Thus, they do not deliver sufficient astrometric accuracy.

Figure~\ref{f:astrometry}a shows the resulting positions of all nearby point sources relative to GJ~758 at E1--E4, as well as the data point for GJ~758~B in E5.  The E4 data points represent the mean positions of the sources in the four datasets that reveal GJ~758~B, i.e.\ the $J$, $H$, $K_\mathrm{c}$, and CH4S data. The error bars for the HiCIAO data (first three epochs) are based on the pixel size of the HiCIAO camera, 9.5\,mas.  For the Gemini data, the pixel size is 22\,mas, yielding an error of $22\,$mas$/\sqrt4 = 11$\,mas for the combined data points.  These errors are consistent with the scatter of the background star data points around the projected motion path, given the expected contributions from the proper motions of the individual background stars.  For the E5 data point, we assume an isotropic error of 10\,mas to represent the anisotropic error bars of 5 and 15\,mas shown in \citet{currie2010}.

We observe that GJ~758~B pursues a trajectory to the northwest consistent with orbital motion relative to GJ~758, clearly setting it apart from the background star trajectory, which is dominated by GJ~758's known parallactic and proper motion.  The source tentatively referred to as ``GJ~758~C'' in P1 is found to follow the background star track; while this was still indistinguishable from orbital motion at the $2\sigma$ level in November 2009, the May 2010 data are unambiguous.  This also allows us to identify the candidate signal in \citet{currie2010} as spurious.

\subsection{Orbital Monte Carlo simulation}

With a total of $N=10$ scalar parameters (2 coordinates $\times$ 5 epochs), the astrometric data on GJ~758~B is now extensive enough to fit synthetic orbital solutions generated by a Monte Carlo simulation with the least-squares method rather than with the simplified approach previously used in P1 and \citet{currie2010}.  However, since the curvature of the orbit is not yet measurable, the benefit of this improvement is limited.

We generate a large number ($>$$10^6$) of orbital trajectories with random values for eccentricity $e$, inclination $i$, argument of periastron $\omega$, and longitude of the ascending node $\Omega$. The distributions are presumed to be flat, except for the inclination, where larger angles are favored proportionately to $\sin i$ in order to represent their higher geometric likelihood.  The two remaining orbital parameters, the semimajor axis $a$ and the mean anomaly at epoch $M_0$, are implicitly chosen by defining an anchor point $(x_\mathrm{A}, y_\mathrm{A})$ in the projected image plane where the companion is located a given epoch $t_\mathrm{A}$.  In order to achieve a high production rate of valid orbital solutions, we choose $t_\mathrm{A}$ to be the mean of the five observational epochs, $\langle t_\mathrm{obs} \rangle$, and generate $(x_\mathrm{A}, y_\mathrm{A})$ randomly in a box of 20\,mas\,$\times$\,20\,mas centered on the mean astrometric coordinates $(\langle x_\mathrm{obs}\rangle, \langle y_\mathrm{obs}\rangle)$.  

In order to evaluate an orbital solution for consistency with the data, we determine the predicted position of the companion at the five epochs and calculate the $\chi^2$ deviation. The minimum best fit achieves $\chi^2=2.09$.  We select the set of orbits with $\chi^2<3.0\approx\chi^2_\mathrm{min}+1$ to represent the ``good fits''.

\begin{figure*}[p]
\centering
\includegraphics[width=0.49\linewidth]{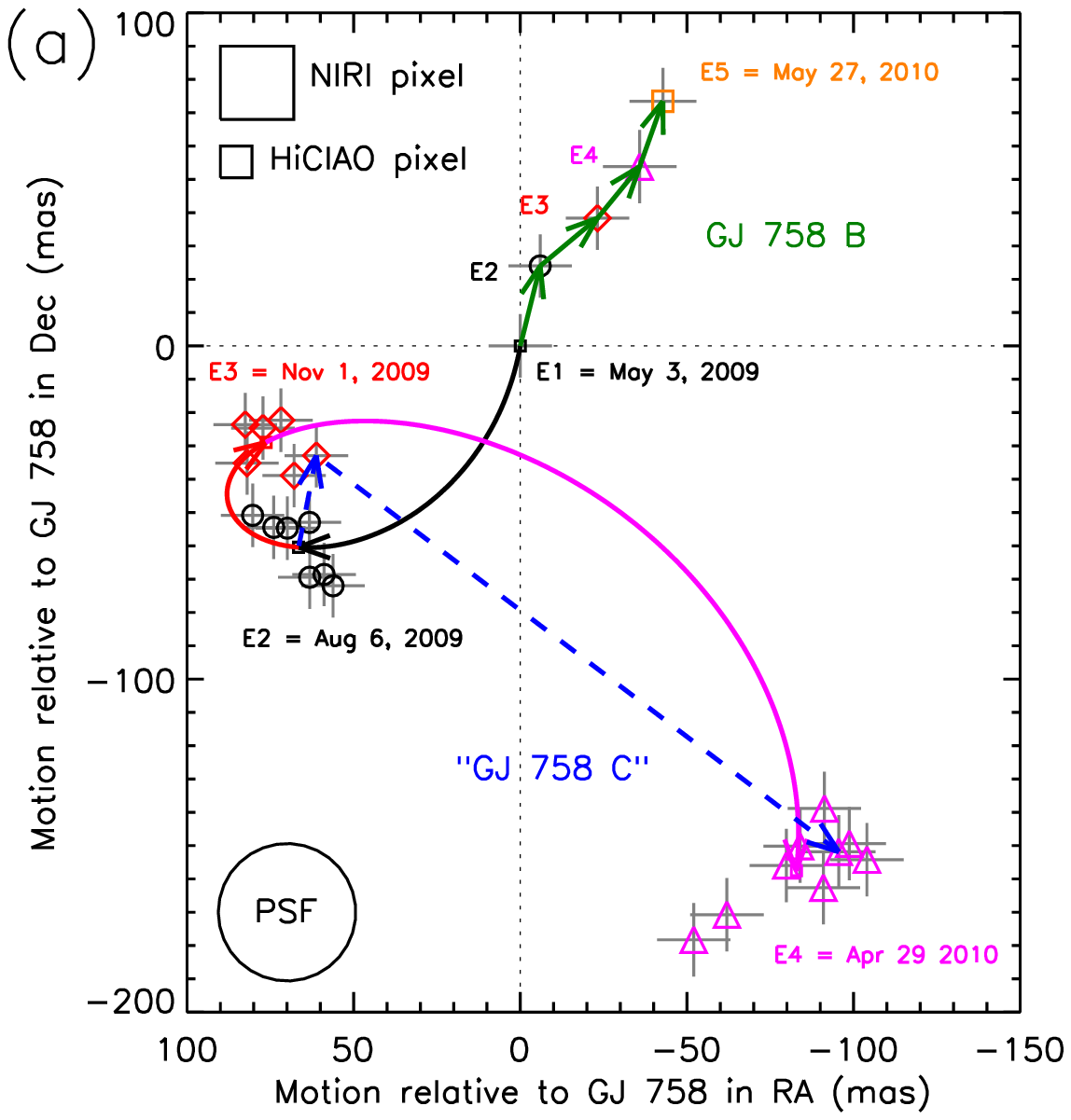}
\includegraphics[width=0.49\linewidth]{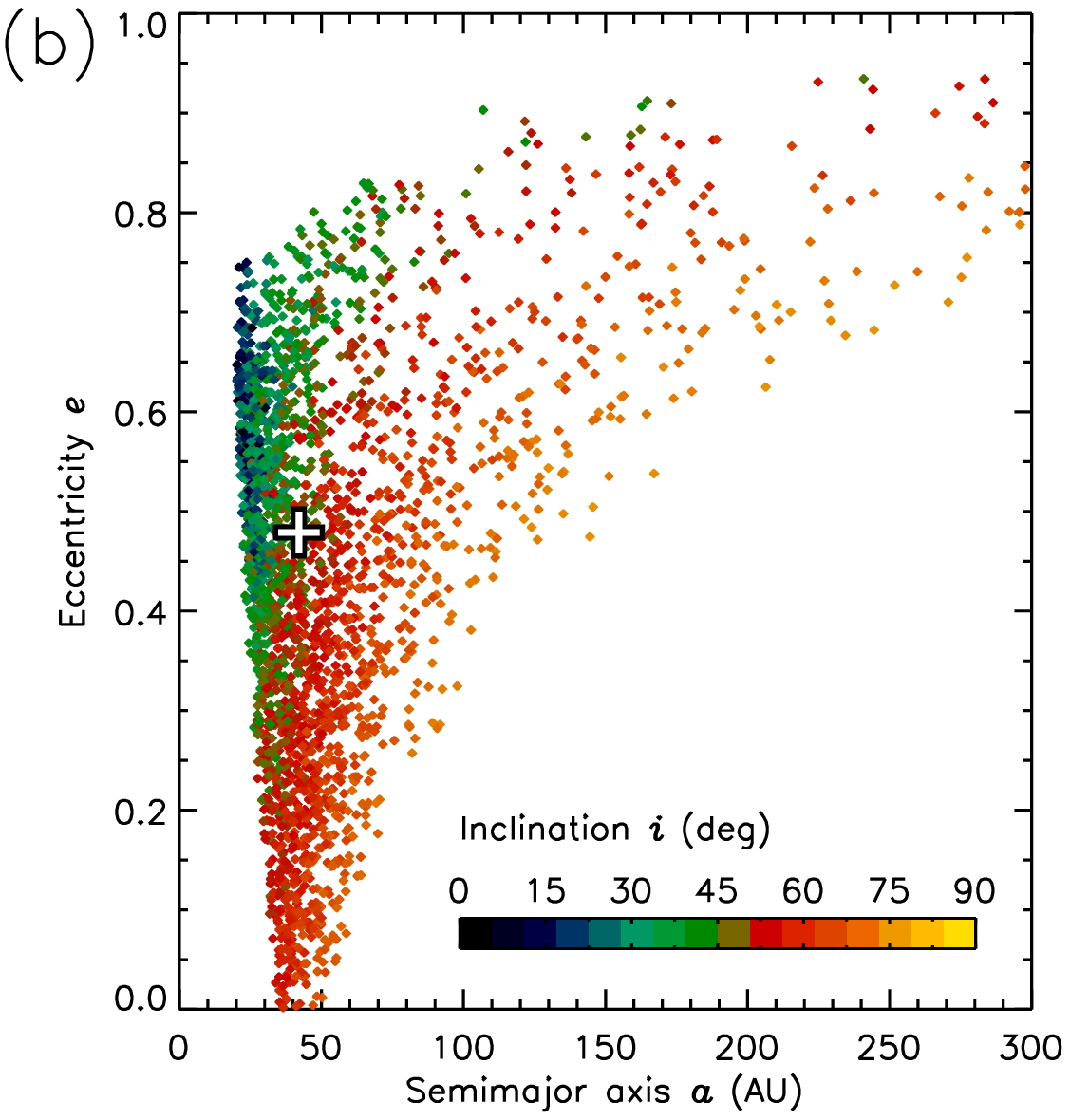}
\caption{Astrometric analysis. (\textbf{a}) Motions of point-sources near GJ~758 across five epochs (E1--E5), measured relative to GJ~758's position. GJ~758~B exhibits common proper motion with its parent star as well as systematic orbital motion towards the northwest, whereas all other point-sources follow the expected trajectory for background stars (solid arrows). The object referred to as ``GJ~758~C'' in P1 is unambiguously identified as a background star (motion highlighted by dashed blue arrows). The grey plus signs are $1\sigma$ error bars. The circle marked as ``PSF'' shows the size of the resolution element in $H$-band on HiCIAO. (\textbf{b}) Plot of eccentricity $e$ against semimajor axis $a$ for 2516 orbital solutions with $\chi^2 \le 3.0$ generated by the Monte Carlo simulation. The orbit selection is biased according to the statistical weight $\langle v\rangle / v_\mathrm{obs}$ as in P1.The inclination $i$ is shown by color coding. The weighted median values of $a$ and $e$ are marked with a white plus sign.  
}
\label{f:astrometry}
\end{figure*}

Figure~\ref{f:astrometry}b illustrates the distribution of semimajor axis $a$, eccentricity $e$ and inclination $i$ for 2516 ``good fit'' orbits.  The solutions appear to lie in a two-dimensional manifold in the three-dimensional parameter space. We note that more stringent fitting requirements (e.g.\ $\chi^2<2.5$) do not relevantly reduce the size of this manifold, showing that this spread is due to model degeneracy rather than fitting errors.  The degeneracy represents the fact that only the projected position of GJ~758~B can be tracked, leaving the line-of-sight component of its position and velocity undetermined. In order to break this degeneracy, it is necessary to measure the curvature of the projected orbit. Given the typical predicted orbital periods of several centuries, this requires monitoring on time scales of at least a decade.

The numerical results are summarized in Table~\ref{t:results}.  We note that the weighted median eccentricity has dropped from 0.73 in \citet{currie2010} to 0.56; this is because the new epochs E3 and E4 suggest that the previous epochs were overestimating the orbital velocity.

\begin{table}[p]
\caption{Numerical Results}
\label{t:results}
\centering

\begin{tabular}{lcc}
\hline
\hline
Photometry (mag) & GJ~758~A & GJ~758~B \\
\hline
$M_{\rm J}$ (1.15-1.33 $\mu$m) & --- & 17.58$\pm$0.20 \\
$M_{\rm H}$ (1.49-1.78 $\mu$m) & --- & 18.16$\pm$0.20 \\
$M_{\rm Kc}$ (2.08-2.11 $\mu$m) & --- & 17.12$\pm$0.20 \\
$M_{\rm L^\prime}$ (3.43-4.13 $\mu$m) & 3.7$\pm$0.1 & 15.0$\pm$0.1 \\
$M_{\rm Ms}$ (4.55-4.79 $\mu$m) & 3.6$\pm$0.1 & $\geq$13.9$^{+0.5}_{-0.8}$ \\
$M_{\rm CH4S}$ (1.53-1.63 $\mu$m) & --- & 17.74$\pm$0.20 \\
$M_{\rm CH4L}$ (1.64-1.74 $\mu$m) & --- & $\geq$18.86$\pm$0.20 \\
\hline
\multicolumn{3}{p{7.5cm}}{}
\end{tabular}

\begin{tabular}{lcc}
\hline
\hline
GJ~758~B orbital parameters & Weighted median & 68\% interval \\
\hline
Semimajor axis $a$ (AU) & 44.8 & 30.7--89.2 \\
Eccentricity $e$ & 0.564 & 0.416--0.712 \\
Inclination $i$ (deg) & 43.6 & 26.2--56.7 \\
Period $P$ (yr) & 299 & 170--843 \\
\hline
\multicolumn{3}{p{7.5cm}}{}
\end{tabular}

\begin{tabular}{lcccc}
\hline\hline
 		&  \multicolumn{4}{c}{GJ~758~B's position relative to GJ~758}\\
Epoch	& $\Delta$RA ($\arcsec$) & $\Delta$Dec ($\arcsec$) & Error ($\arcsec$) 
	& Ref.\\
\hline
E1: May 3, 2009 & $-0.574$ & $-1.789$ & 0.010 & (1)\\
E2: Aug 8, 2009 & $-0.579$ & $-1.765$ & 0.010 & (1)\\
E3: Nov 1, 2009 & $-0.597$ & $-1.751$ & 0.010 & (2)\\
E4: Apr 29, 2010 & $-0.609$ & $-1.735$ & 0.011 & (2)\\
E5: May 27, 2010 & $-0.616$ & $-1.716$ & 0.010 & (3)\\
E6: Aug 6, 2010 & \multicolumn{4}{c}{ --- insufficient astrometry --- }\\
\hline
\multicolumn{5}{p{7.5cm}}{
	\textbf{References.} (1) P1, (2) this work, (3) \citet{currie2010}.  Note that the error bars for astrometry in E1 and E2 were mistakenly listed as 5\,mas in Table~1 in (1); however, the text and plots use the correct value of 9.5\,mas.
}\\
\hline
\end{tabular}
\end{table}

\acknowledgements
We thank David Lafreni\`ere for providing us with the source code for LOCI, Eric Mamajek for useful discussion, the staff at Subaru, Gemini and Keck for their support, and an anonymous referee for useful suggestions. The Gemini time was allocated by NOAO. The Keck telescope was funded by the W.M. Keck foundation, and time was allocated by NASA through partnership with Caltech and UC. We acknowledge the cultural significance that the summit of Mauna Kea has to the indigenous Hawaiian community. Part of the research was supported by NSF grants AST 1009203, AST 0802230, AST 1009314, the AAS Chretien grant, a Grant-in-Aid for Specially Promoted Research, the Mitsubishi Foundation, and JPL, Caltech under NASA contract.

{\it Facilities:} \facility{Subaru (HiCIAO, AO188)}, 
\facility{Gemini North (NIRI)}, \facility{Keck (NIRC2)}.

\clearpage


\clearpage

\end{document}